# Compressive Sensing Based Opportunistic Protocol for Throughput Improvement in Wireless Networks

Syed T. Qaseem[1], and Tareq Y. Al-Naffouri[2]

[1]LCC, Riyadh, Saudi Arabia

[2]King Fahd University of Petroleum & Minerals, Dhahran, Saudi Arabia

**Abstract**

A key feature in the design of any MAC protocol is the throughput it can provide. In wireless networks, the channel of a user is not fixed but varies randomly. Thus, in order to maximize the throughput of the MAC protocol at any given time, only users with large channel gains should be allowed to transmit. In this paper, compressive sensing based opportunistic protocol for throughput improvement in wireless networks is proposed. The protocol is based on the traditional protocol of R-ALOHA which allows users to compete for channel access before reserving the channel to the best user. We use compressive sensing to find the best user, and show that the proposed protocol requires less time for reservation and so it outperforms other schemes proposed in literature. This makes the protocol particularly suitable for enhancing R-ALOHA in fast fading environments. We consider both analog and digital versions of the protocol where the channel gains sent by the user are analog and digital, respectively.

## I. INTRODUCTION

The topic of multiple access schemes has drawn the attention of researchers and developers with every new technology in communications [1]. For wireless networks, random access protocols (ALOHA, slotted ALOHA, ... etc.) are very popular as they use the shared medium for transmission with no coordination required among participants. This however results in data packet collisions thereby reducing the throughput. In order to reduce these collisions, reservation ALOHA (R-ALOHA) was introduced where the transmission time is divided into frames which are further divided into slots. Within each frame, the first few slots are used for reserving the frame for the best user for data transmission [2].

Recently, opportunistic or channel-aware ALOHA was proposed to improve random access protocols by exploiting *multiuser diversity* [1]. In this situation the wireless channel is accessed by users who have the best channels and hence can support a higher transmission rate. The process of exploiting multiuser diversity can be either centralized or distributed.

In the centralized approach [3], the transmission decision (i.e., which user to transmit data packets) is made by the base station or access node and requires knowledge of each user's channel gain. Major disadvantage of such approach is that the time required to measure all the users' channels grows linearly with the number of users [4], as users transmit their channel related information orthogonally in time. Therefore, for systems with large number of users, this may not be possible as the overhead (time required to measure the channels) may exceed the coherence time of the system.

Contrast this with the distributed approach [1]. Here, transmission decisions are individually made by each user based on its local channel information. Thus, in reciprocal networks (as in time-division duplex systems[1]) the base station broadcasts a pilot signal to all users, and each user measures its own channel gain using this pilot signal and makes its transmission decision accordingly. Note that the overhead in this approach is independent of the number of users.

Combining reservation and opportunistic communication, Qin & Berry proposed in [4]- [5] a distributed algorithm for random access where *strong* users (users whose channel gains are above a certain threshold) send reservation packets containing their user ID and channel quality information (CQI)[2] and collisions in the reservation phase are resolved via splitting. A splitting algorithm uses some tree-like mechanism where users involved in a collision are divided into several subsets and only the user or users in one of the subsets transmits at the next time slot so as to reduce the probability of collision [2], [4]- [5]. Thus, the splitting algorithm resolves a collision which eventually results in finding the user with the best channel gain out of all backlogged users. As the frame duration is limited by the coherence time of the channel, reducing the reservation time will increase the time available for the best user to transmit data, thereby increasing the throughput. Qin and Berry showed that there scheme requires only 2.5 slots[3]

[1]In a time-division duplex (TDD) system, both uplink and downlink use the same carrier frequency, and therefore, downlink channel estimates can be used for the uplink too.

[2]We use the terms "CQI" and "channel gain" interchangeably throughout the rest of the paper.

[3]In [4]- [5], frames and slots are refereed as slots and mini-slots, respectively.

(*on average*) for reserving the frame for the best user and so the throughput will improve for channels having larger coherence time [4]- [5]. To the best of our knowledge, this is the best result in R-ALOHA literature.

Despite the low average number of slots required to reserve the best user, two major drawbacks associated with this scheme are i) the length of a slot must be greater than the round-trip time (RTT), and ii) channel coherence time (CCT) must be large in order to achieve near-optimal performance [4]- [5]. It is important to note that there is little that one can do to reduce RTT or increase CCT as these are the fundamental limitations of any practical wireless network. This is because CCT depends on the physical environment and RTT is limited by the cell size.

In this paper, we consider a distributed approach based random access for the uplink. Here also within each frame, we reserve the first few slots for sending reservation packets. However, in our case the number of reservation slots is fixed and is decided beforehand. All users whose CQI is above a particular threshold contend for reservation by simultaneously reporting their CQIs after multiplying them with a unique random binary chip sequence, and remain silent otherwise. We use the emerging compressive sensing (CS) technique to identify users who have fed back and to estimate the fedback CQIs.

From the reservation standpoint, our scheme differs from [4]- [5] in the following:

1) Slot duration ($T_s$) is not limited by the RTT.

2) No user ID is fedback.

3) There is a fixed number of reservation slots. This makes the algorithm simpler as the intelligence needed by higher layers is minimized.

The remainder of the paper is organized as follows. In Section II, the system model is introduced. In Section III we discuss the proposed scheme and study its throughput in Section IV. In Section V, we evaluate the performance of the proposed scheme. This is followed by the numerical results and conclusions in Sections VI and VII, respectively.

## II. System Model

### A. Channel Model

We consider a TDD system and R-ALOHA based multi-access model where all $n$ users are backlogged and always have data to send. The total time is divided into frames each with duration $T_f$ equal to channel coherence time $T_c$. Note that the frames are further divided into $p$ slots of duration $T_s$, i.e. $T_f = T_c = pT_s$. Also, each uplink frame of duration $T_c$ is divided into two parts: reservation phase of duration $T_r = mT_s$ for selecting the best user, followed by data transmission phase of duration $T_d = (p - m)T_s$ for the selected user to send data. Pictorial representation of the frame structure is shown in Fig. 1. The details of the figure will be discussed later in the paper. We assume that at the start of each frame, each user has knowledge of its own channel gain, but not the gain of any of the other users[4]. Furthermore, we assume that the users experience i.i.d. Rayleigh fading.

## B. Contention Model

We present here a compressive sensing model for reserving the frame to the best user i.e, the user with the best CQI. In order to exploit multiuser diversity, the number of slots $r$ reserved for reporting CQI of strong users to the base station (BS) are fixed and shared among all $n$ users. In the analog case, each strong user multiples its CQI (denoted as $h_i$) with a random binary chip sequence (consisting of $\pm 1$ each with probability 0.5)[5] of length equal to $r$ slots ($rT_s$) before sending it over the multiple access shared channels. Thus, if user $i$ is strong, it will multiply $v_i = h_i$ with a random binary sequence (column vector) $\mathbf{a}_i$. If user $i$ is a weak user, it remains silent or effectively sends $v_i = 0$ multiplied by its random binary sequence $\mathbf{a}_i$. In the digital case, the same procedure is adopted except that we set $v_i = 1$ for the strong user. In a nutshell, the model can be described as

$$\begin{bmatrix} y_1 \\ y_2 \\ \vdots \\ y_r \end{bmatrix} = \begin{bmatrix} \mathbf{a}_1 & \mathbf{a}_2 & \cdots & \mathbf{a}_n \end{bmatrix} \begin{bmatrix} v_1 \\ v_2 \\ \vdots \\ v_n \end{bmatrix}$$

or

$$\mathbf{y} = \mathbf{A}\mathbf{v} \tag{1}$$

where $\mathbf{A}$ is a $r \times n$ Bernoulli matrix [6] with $r \ll n$ and where $\mathbf{v}$ is a sparse vector with $s$ non zero entries $\|\mathbf{v}\|_0 = s$, where $\| \cdot \|_0$ is the combinatorial norm $\ell_0$. Thus, in the analog case,

$$v_i = \begin{cases} h_i & \text{if } h_i \geq \zeta; \\ 0 & \text{if } h_i < \zeta. \end{cases}$$

---

[4]In reciprocal networks (as in TDD) this knowledge could be gained by having the base station broadcast a pilot signal at the start of each frame.

[5]There are two ways of assigning chip sequences to the users: pre-programmed in users' device or by sending it over the air.

[6]CS can be applied as Bernoulli matrices are shown to satisfy the RIP [7].

whereas in the digital case

$$v_i = \begin{cases} 1 & \text{if } h_i \geq \zeta; \\ 0 & \text{if } h_i < \zeta. \end{cases}$$

where $\zeta$ is the threshold that distinguishes weak users from strong ones.

The BS receives the users' requests and finds the strong users via compressive sensing. The threshold $\zeta$ is set according to the desired sparsity level $s$ and the strongest user among them is selected. To reduce the number of users $s$ who contend, we pursue a thresholding strategy where the user will send reservation packets if his CQI is greater than a threshold $\zeta$ to be determined. Noting that the users' CQI are i.i.d., we can choose $\zeta$ to produce a sparsity level $s$. This happens provided that

$$\bar{F}_h(\zeta) = \arg \max_{u \in (0,1)} \binom{n}{s} u^s (1-u)^{n-s} \tag{2}$$

where $\bar{F}_h(\zeta)$ is the complementary cumulative distribution function (CCDF) of the channel gain ($h_i$'s) defined as $\bar{F}_h(\zeta) = \mathbb{P}[h_i > \zeta] = e^{-\zeta}, \zeta \geq 0$. The threshold that maximizes (2) can be calculated as follows. Let $\psi = \binom{n}{s} u^s (1-u)^{n-s}$. Differentiating $\psi$ w.r.t $u$ and setting the derivative to 0, and solving for $u$ yields $u = s/n$. Thus, $\bar{F}_h(\zeta) = s/n$, or

$$\zeta = \bar{F}_h^{-1} \left( \frac{s}{n} \right) \tag{3}$$

## III. The Proposed Protocol

Before we discuss the proposed protocols, we present an overview of compressive sensing results related to our work.

### A. Compressive Sensing Overview

Compressive sensing refers to the recovery of sparse signals from limited measurements. Approaches for recovering the sparsity pattern $S$ with $|S| = \|\mathbf{v}\|_0 = s$ of signal $\mathbf{v} \in \mathbb{R}^n$, i.e. $S = \{i \in \{1, \ldots, n\} | v_i \neq 0\}$, in the setting (cf. (1)) are discussed here.

*1) Sparsity Pattern Recovery Results:* The sparsest solution to underdetermined systems of linear equations is given by

$$\min_{\mathbf{v} \in \mathbb{R}^n} \|\mathbf{v}\|_0 \qquad \text{s. t.} \qquad \mathbf{y} = \mathbf{A}\mathbf{v} \tag{4}$$

The solution to this problem is in general NP hard [6]. This computational intractability has recently led researchers to develop alternatives to (4). Recently, $\ell_1$-minimization (Basis Pursuit) has been proposed as a convex alternative to the $\ell_0$ [6]

$$\min_{\mathbf{v} \in \mathbb{R}^n} \|\mathbf{v}\|_1 \qquad \text{s. t.} \qquad \mathbf{y} = \mathbf{A}\mathbf{v}. \tag{5}$$

A recent paper by Candes, Romberg & Tao [7] shows that when $\mathbf{A}$ is a random matrix with i.i.d. entries from a suitable distribution, all sparse signals $\mathbf{v}$ with sparsity level $s$ can be recovered using (5) with very high probability provided the number of measurements (or channels) satisfy $r \sim O(s \log(n/s)) = cs \log(n/s)$, where $c$ is a constant. Similar results are obtained using matching pursuit algorithms e.g., maximum correlation [8] and CoSaMP [9], which are simpler methods compared to $\ell_1$-minimization.

*2) Estimating or Refining Sparse Signal:* Once the sparsity pattern $S$ is determined, least squares can be used to estimate or refine the active elements of $\mathbf{v}$ as follows [10]:

$$\mathbf{v}_S^{ls} = (\mathbf{A}_S^* \mathbf{A}_S)^{-1} \mathbf{A}_S^* \mathbf{y} \qquad (6)$$

where $\mathbf{A}_S$ denotes the sub-matrix formed by the columns $\{\mathbf{a}_j : j \in S\}$, indexed by the sparsity pattern $S$.

### B. Compressive Sensing based Reservation Protocols

Here, we present the compressive sensing based protocols for reserving the frame for the best user. We consider both analog and digital cases depending on whether the CQI sent by the users for reservation are analog or digital. The proposed protocols are as follows:

*1) Analog Case:* In the analog case, users whose CQI is above a certain threshold $\zeta$ as determined by (3) send their analog CQI value, and remain silent otherwise. Thus the vector $\mathbf{v}$ in (1) is sparse with sparsity level determined by the number of users who send their CQIs. The BS receives the users' requests and finds the strong users via compressive sensing. The reservation procedure is thus as follows:

1) **Threshold Determination:** The BS decides on thresholding level $\zeta$ based on the sparsity level that can be recovered with the aid of (3).

2) **Contention & Contention Resolution:**

   - CQI Determination: Each user determines his CQI.
   - Reservation: Each user whose CQI is above the threshold $\zeta$ send a reservation packet according to input/output equation (1). Otherwise, the user remains silent.
   - Compressive Sensing: BS finds the strong users using Compressive Sensing.
   - Least-squares estimation/refining: BS estimates or refines results obtained via CS using least-squares (6).

3) **User Selection:** BS selects the strongest user among the strong users.

*2) Digital Case:* The digital case is similar to analog case except that each user sends "1" if its CQI is above a particular threshold. Otherwise, the user remains silent. To increase the reservation granularity, we let the users compare their CQI to a set of thresholds, not just one. Thus, suppose that we want to set $k$ thresholds $\zeta_1 < \zeta_2 < ...... < \zeta_k$ such that the number of users whose CQI lie between the two consecutive thresholds $Q_i = [\zeta_i, \zeta_{i+1})$ is equal to $s$. Note that the last interval is $[\zeta_k, \infty)$ as $\zeta_{k+1} = \infty$. Using (3), we can set the lowermost threshold as

$$\bar{F}_h(\zeta_1)n = sk, \quad \text{or,} \quad \zeta_1 = \bar{F}_h^{-1}\left(\frac{sk}{n}\right) \tag{7}$$

Continuing in the same way, we get

$$\zeta_2 = \bar{F}_h^{-1}\left(\frac{s(k-1)}{n}\right), \cdots\cdots, \zeta_k = \bar{F}_h^{-1}\left(\frac{s}{n}\right). \tag{8}$$

The reservation procedure is thus as follows:

1) **Threshold Determination:** The BS decides on thresholding levels $\zeta_1, \zeta_2, ..., \zeta_k$ based on the sparsity level that can be recovered with the aid of (7) & (8).

2) **Contention & Contention Resolution:** Repeat the following steps for each threshold interval $[\zeta_i, \zeta_{i+1})$, $i = 1, \cdots, k$:

   - CQI Determination: Each user determines his CQI.
   - Reservation: Each user whose CQI lies in the threshold interval $[\zeta_i, \zeta_{i+1})$, sends a reservation packet according to input/output equation (1). Otherwise, the user remains silent.
   - Compressive Sensing: BS finds the strong users using Compressive Sensing.

3) **User Selection:** BS randomly selects[7] one of strong users of the highest active threshold interval. The threshold interval is considered active if there is at least one user sending a reservation packet in that interval. Thus, it is evident that with more number of threshold levels, higher accuracy of CQI will be achieved. Note that the rate at which data will be transmitted is determined by the lower limit of the highest active threshold interval.

## IV. THROUGHPUT

In this section, we study throughput achieved by the proposed protocol for both analog and digital cases. From Fig. 1, we see that the portion of the frame used for sending data is $\left(1 - \frac{T_r}{T_c}\right)$. Thus, the

---

[7]Here threshold should be set such that there is only one strong user ($s = 1$) who sends reservation packet. This is because within the threshold interval multiuser diversity can't be exploited as users' CQI within any threshold interval are represented as "1". However because of the random nature of the channel there may be multiple users who are strong. In this case, the BS randomly selects one of the strong users.

throughput achieved by the proposed scheme is given by

$$\mathcal{C} = \left(1 - \frac{T_r}{T_c}\right)\mathcal{R} = \left(1 - \frac{m}{p}\right)\mathcal{R}$$

where $\mathcal{R}$ is the maximum possible rate at which data can be transmitted.

### A. Analog Case

In the analog case, the throughput $\mathcal{R}_a$ is given by

$$\mathcal{R}_a = \mathbb{E}\left[\log_2(1 + \max_{1 \le i \le s} h_i) \mid \mathcal{A}, \mathcal{B}\right]\mathbb{P}(\mathcal{A} \cap \mathcal{B}) \qquad (9)$$

where $\mathbb{P}(\cdot)$ is the probability of an event, $\mathcal{A}$ is the event that CS is successful, and $\mathcal{B}$ is the event that $\max_{1 \le i \le s} h_i > \zeta$ is not a null set. As the events $\mathcal{A}$ and $\mathcal{B}$ are independent, we can write (9) as

$$\mathcal{R}_a = \mathbb{E}\left[\log_2(1 + \max_{1 \le i \le s} h_i) \mid \mathcal{A}, \mathcal{B}\right]\mathbb{P}(\mathcal{A})\mathbb{P}(\mathcal{B}). \qquad (10)$$

We assume that given a fixed number of reservation slots $r$, the sparsity level $s$ is chosen low enough (by appropriately choosing $\zeta$) such that $\mathbb{P}(\mathcal{A}) \to 1$. From compressive sensing theory we know that given $r$ reservation slots, the maximum sparsity level $s$ that can be handled successfully is related to $r$ by $r = c\log(s/n)$. Thus, the threshold $\zeta$ should be set such that only $s$ out of $n$ users contend for reservation. Using (3), this is achieved by setting $\zeta = -\log(s/n)$ as $\bar{F}^{-1} = -\log(\cdot)$ for a Rayleigh fading channel. Thus,

$$\mathcal{R}_a = \mathbb{E}\left[\log_2(1 + \max_{1 \le i \le s} h_i) \mid \mathcal{B}\right]\mathbb{P}(\mathcal{B}) \qquad (11)$$

$$= \mathbb{E}\left[\log_2(1 + \max_{1 \le i \le s} h_i) \mid \mathcal{B}\right](1 - \mathbb{P}(\mathcal{B}^{\mathcal{C}})) \qquad (12)$$

Note that $\mathbb{P}(\mathcal{B}) = (1 - \mathbb{P}(\mathcal{B}^{\mathcal{C}}))$, where $\mathcal{B}^{\mathcal{C}}$ is the event that the CQIs of all users are below the threshold $\zeta$. The probability of this event is given by

$$\mathbb{P}(\mathcal{B}^{\mathcal{C}}) = \mathbb{P}(\max_{1 \le i \le n} h_i \le \zeta) = [F_h(\zeta)]^n = (1 - e^{-\zeta})^n = \left(1 - \frac{s}{n}\right)^n$$

where $F_h(\zeta)$ is the cumulative distribution function (CDF) of $h_i$, and $F_h(\zeta) = (1 - e^{-\zeta})$ for Rayleigh fading channel. Note that $\mathbb{P}(\mathcal{B}^{\mathcal{C}}) \to 0$ for large $n$, which implies that for systems with large number of users, the hit on the throughput due to thresholding is negligible. We are left to evaluate $\mathbb{E}\left[\log_2(1 + \max_{1 \le i \le n} h_i) \mid \mathcal{B}\right]$ which can be written as

$$\mathbb{E}\left[\log_2(1 + \max_{1 \le i \le s} h_i) \mid \mathcal{B}\right] = \int_\zeta^\infty \log_2(1 + x)f_\gamma(x)dx \qquad (13)$$

where $\gamma = \max_{1 \leq i \leq s} h_i$, $h_i \geq \zeta$. In order to evaluate this expression, we need to find the PDF of $\gamma$, $f_\gamma(x)$. The rest of this subsection is devoted to finding this PDF.

We can write the conditional CDF of $h_i$ given $h_i \geq \zeta$ as

$$\mathbb{P}(h_i < x | h_i \geq \zeta) = \begin{cases} \frac{F_h(x) - F_h(\zeta)}{1 - F_h(\zeta)} & \text{if } x \geq \zeta; \\ 0 & \text{if } x < \zeta. \end{cases}$$

Thus,

$$\mathbb{P}\left[\max_{1 \leq i \leq s}(h_i < x | h_i \geq \zeta)\right] = \left[\frac{F_h(x) - F_h(\zeta)}{1 - F_h(\zeta)}\right]^s \tag{14}$$

The PDF, $f_\gamma(x)$, is obtained by differentiating the above expression to get

$$f_\gamma(x) = s \left(\frac{F_h(x) - F_h(\zeta)}{1 - F_h(\zeta)}\right)^{(s-1)} \cdot \left(\frac{f_h(x)}{1 - F_h(\zeta)}\right) \tag{15}$$

$$= \frac{s f_h(x) \left[F_h(x) - F_h(\zeta)\right]^{(s-1)}}{\left[1 - F_h(\zeta)\right]^s} \tag{16}$$

where $f_h(x) = e^{-x}$ and $F_h(x) = 1 - e^{-x}$ which follows from the fact the users channels are Rayleigh fading. Now substituting $\zeta = -\log(s/n)$ in (16) yields

$$f_\gamma(x) = \frac{s e^{-x} \left(\frac{s}{n} - e^{-x}\right)^{(s-1)}}{\left(\frac{s}{n}\right)^s} \tag{17}$$

which allows us to write (13) as

$$\mathbb{E}\left[\log_2(1 + \max_{1 \leq i \leq s} h_i) \mid \mathcal{B}\right] = \frac{n^s}{s^{s-1}} \int_\zeta^\infty \log_2(1 + x) e^{-x} \left(\frac{s}{n} - e^{-x}\right)^{(s-1)} dx \tag{18}$$

Thus,

$$\mathcal{R}_a = \left(1 - \left(1 - \frac{s}{n}\right)^n\right) \frac{n^s}{s^{s-1}} \int_\zeta^\infty \log_2(1 + x) e^{-x} \left(\frac{s}{n} - e^{-x}\right)^{(s-1)} dx \tag{19}$$

### B. Analog Case: Asymptotic Analysis

To get a better understanding of the throughput achieved in the analog case, let's study the asymptotic case of large number of users. Starting from (12) and using the approach given in [11], we can write the throughput as

$$\mathcal{R}_a = \mathbb{E}\left[\log_2(1 + \max_{1 \leq i \leq s} h_i) \mid \mathcal{B}\right] (1 - \mathbb{P}(\mathcal{B}^{\mathcal{C}})) \tag{20}$$

$$\geq \log_2(1 + \zeta)(1 - \mathbb{P}(\mathcal{B}^{\mathcal{C}})) \tag{21}$$

where $\mathbb{P}(\mathcal{B}^C)$ can be written as

$$\mathbb{P}(\mathcal{B}^C) \stackrel{(a)}{=} (1-\eta)^n$$

$$\stackrel{(b)}{\leq} e^{-\eta m}$$

Here, $(a)$ follows because $F_h(\zeta) = \mathbb{P}[h_i \leq \zeta] = 1 - e^{-\zeta}$, so that $\eta = e^{-\zeta}$, and $(b)$ follows because $(1-\eta)^{-1} \geq e^\eta$ for $|\eta| < 1$. This allows us to write (21) as

$$\mathcal{R}_a \geq \log_2(1+\zeta)(1-e^{-\eta m})$$

$$= \log_2(1 - \log(s/n))(1 - e^{-s})$$

where the last line follows by substituting, $\zeta = -\log(s/n)$. Note that $\eta = e^{-\zeta}$, so we get $\eta = s/n$. Now by setting $s = \log(n)$, we get $(1 - e^{-s}) = \left(1 - \frac{1}{n}\right)$ and therefore,

$$\mathcal{R}_a \geq \log_2(1 + \log(n) - \log\log(n)) \left(1 - \frac{1}{n}\right) \tag{22}$$

As $\lim_{n\to\infty} \frac{\log\log(n)}{\log(n)} \to 0$, so form the above equation we deduce that $\lim_{n\to\infty} \frac{\mathcal{R}_{oc}}{\mathcal{R}_a} = 1$, where $\mathcal{R}_{oc} = \log_2(1 + \log(n))$ is the optimal rate of the centralized scheduling scheme [3]. Thus,

$$\mathcal{C} = \left(1 - \frac{T_r}{T_c}\right)\mathcal{R}_{oc} = \left(1 - \frac{m}{p}\right)\mathcal{R}_{oc}. \tag{23}$$

Note that we have proved that the only hit on the throughput is due to portion of the frame used for reservation. However, in order to achieve this optimal transmission rate, the following conditions should be met:

1) the number of users who contend for reservation $s$ must scale as $\log(n)$.

2) as a consequence of 1), the number of reservation slots must scale as $r = cs\log(n/s) = c((\log(n))^2 - (\log(n)\log\log(n)))$.

## C. Digital Case

In the digital case, the throughput $\mathcal{R}_d$ is given by

$$\mathcal{R}_d = \mathbb{E}\left[\log_2(1 + \max_{1 \leq i \leq k} \zeta_i) \mid \mathcal{A}, \mathcal{B}\right]\mathbb{P}(\mathcal{A})\mathbb{P}(\mathcal{B}) \tag{24}$$

as $\mathcal{A}$ and $\mathcal{B}$ are independent events, where $\mathcal{A}$ is the event that CS is successful and $\mathcal{B}$ is the event that $\max_{1 \leq i \leq k} \zeta_i$ is not a null set which basically means $\max_{1 \leq i \leq n} h_i > \zeta_1$ is not a null set. Note that $\max_{1 \leq i \leq k} \zeta_i$ is the lower limit of the highest active threshold interval.

Now choosing $r = O(s \log(n/s))$ results in $\mathbb{P}(\mathcal{A}) \to 1$ exponentially with $r$. Thus, we can write (24) as

$$\mathcal{R}_d = \mathbb{E}\left[\log_2(1 + \max_{1 \le i \le k} \zeta_i) \mid \mathcal{B}\right] \mathbb{P}(\mathcal{B}) \tag{25}$$

$$= \mathbb{E}\left[\log_2(1 + \max_{1 \le i \le k} \zeta_i) \mid \mathcal{B}\right] (1 - \mathbb{P}(\mathcal{B}^{\mathcal{C}})) \tag{26}$$

where $\mathbb{P}(\mathcal{B}^{\mathcal{C}})$ can be calculated as

$$\mathbb{P}(\mathcal{B}^{\mathcal{C}}) = \mathbb{P}(\max_{1 \le i \le n} h_i \le \zeta_1)$$

$$\stackrel{(a)}{=} (1 - e^{-\zeta_1})^n$$

$$\stackrel{(b)}{=} \left(1 - \frac{sk}{n}\right)^n$$

where $(a)$ follows because $F_h(\zeta_1) = \mathbb{P}[h_i \le \zeta_1] = 1 - e^{-\zeta_1}$, and $(b)$ follows from (7) and the fact that $\zeta_1 = -\log(sk/n)$. Note that $\mathbb{P}(\mathcal{B}^{\mathcal{C}}) \to 0$ for large $n$, which implies that for systems with large number of users, the hit on the throughput due to thresholding is negligible.

Thus,

$$\mathcal{R}_d = \mathbb{E}\left[\log_2(1 + \max_{1 \le i \le k} \zeta_i) \mid \mathcal{B}\right] \left(1 - \left(1 - \frac{sk}{n}\right)^n\right). \tag{27}$$

We are left to evaluate $\mathbb{E}\left[\log_2(1 + \max_{1 \le i \le k} \zeta_i) \mid \mathcal{B}\right]$ which can be derived analytically as [12]

$$\mathbb{E}\left[\log_2(1 + \max_{1 \le i \le k} \zeta_i) \mid \mathcal{B}\right] = \sum_{i=1}^{k} \log_2(1 + \zeta_i) \mathbb{P}(\text{selected user is in } Q_i) \mathbb{P}(Q_i)$$

where the probability of the threshold interval $Q_i$ is given by $\mathbb{P}(Q_i) = [F_h(\zeta_{i+1}) - F_h(\zeta_i)]$, and where the probability that selected user is in the threshold interval $Q_i$ is given as

$$\mathbb{P}(\text{selected user is in } Q_i) = \sum_{j=0}^{n-1} \frac{1}{j+1} \binom{n-1}{j} \mathcal{P}_1 \mathcal{P}_2.$$

Here,

$$\mathcal{P}_1 = \mathbb{P}(j \text{ users other than the selected user are in } Q_i) = [F_h(\zeta_{i+1}) - F_h(\zeta_i)]^j, \text{ and}$$

$$\mathcal{P}_2 = \mathbb{P}((n-j-1) \text{ users lies below the interval } Q_i) = [F_h(\zeta_i)]^{(n-j-1)}.$$

Substituting these values of $\mathcal{P}_1$ and $\mathcal{P}_2$ yields after some manipulations

$$\mathbb{P}(\text{selected user is in } Q_i) = \frac{[F_h(\zeta_{i+1})]^n - [F_h(\zeta_i)]^n}{[F_h(\zeta_{i+1}) - F_h(\zeta_i)]}.$$

Thus,

$$\mathbb{E}\left[\log_2(1 + \max_{1 \le i \le k} \zeta_i) \mid \mathcal{B}\right] = \sum_{i=1}^{k} \log_2(1 + \zeta_i)([F_h(\zeta_{i+1})]^n - [F_h(\zeta_i)]^n), \tag{28}$$

so that

$$\mathcal{R}_d \quad = (1 - [F_h(\zeta_1)]^n) \sum_{i=1}^{k} \log_2(1 + \zeta_i)([F_h(\zeta_{i+1})]^n - [F_h(\zeta_i)]^n). \tag{29}$$

## V. Performance Evaluation: Reservation time ($T_r$)

In this section, we compare the reservation time required for our scheme with that of Qin & Berry [4]-[5]. Note that reduction in the reservation time allows more time for data transmission, thereby improving the throughput.

The scheme proposed in [4]- [5] (see Fig. 2) requires each strong user to send a reservation packet (containing its ID and CQI) in each slot to the BS and then to wait for the base station to tell whether the slot was idle, contained a successful transmission, or contained a collision. In the case of unsuccessful transmission, splitting is done and this process is continued until the best user is found or there are no more slots in the frame. Thus,

$$T_r = \beta T_s' \tag{30}$$

where $\beta$ is the number of slots required to find the best user, and $T_s'$ is the duration of a slot in [4]- [5]. Note that $\beta \geq 1$ with average value approximately equal to 2.5.

However, in our scheme, strong users send their reservation packets to the BS where it uses compressive sensing to find the best user and then informs the selected user of its decision as shown in Fig. 1. So the base station communicates only once with the user during the reservation time. Moreover, in our scheme, the user does not send its ID as CS is used to determine which users were active and the value of the corresponding CQI. Thus,

$$T_r = T_{CS} + \text{RTT} = mT_s \tag{31}$$

where $T_{CS}$ is the reservation time corresponding to slots required for CS, and $T_s$ is the duration of a slot in our scheme. Note that RTT can be written as $\text{RTT} = tT_s$, where $t$ depends on $T_s$ as RTT is fixed.

### A. Analog Case

In the analog version of the proposed scheme, $T_s = T_a$, where $T_a$ is the time required to transmit one real number. Thus, the time required for reserving the frame for the best user for the scheme proposed in this paper is given by

$$T_r = rT_a + \text{RTT} = (r + t)T_a \tag{32}$$

where $\text{RTT} = tT_a$, and $r = cs \log(n/s)$.

However, in [4]- [5] each slot is comprised of the time required to transmit two real numbers (one for the user ID and the other for CQI), and the RTT. Thus, the time required for reserving the frame for the best user for Qin & Berry's scheme is given by

$$T_r = \beta T_s' = \beta(\text{RTT} + 2T_a) = \beta(t+2)T_a. \tag{33}$$

Thus, our scheme is relatively more efficient than that of Qin & Berry when

$$\beta(t+2)T_a > (t+r)T_a \tag{34}$$

$$\text{i.e.,} \qquad \beta > \frac{t+r}{t+2}. \tag{35}$$

Note that (35) applies to the case when $r \geq 2$, as $\beta \geq 1$. For the case when $r < 2$, our scheme is always better.

### B. Digital Case

In the digital version of the proposed scheme, $T_s = T_b$, where $T_b$ is the time required to transmit one bit. Thus, the time required for reserving the frame for the best user for the scheme proposed in this paper is given by

$$T_r = krT_b + \text{RTT} = (kr + t')T_b \tag{36}$$

where $\text{RTT} = t'T_b$, and $kr$ is the total number of bits required for finding the best user using CS. This is because there are $k$ thresholds, and for each threshold $r$ slots are needed by CS. Also, note that $r = c\log(n)$, as we set $s = 1$ for each threshold interval.

However, in [4]- [5] for the digital case, $\log_2(n)$ bits are required for unique representation of users as there are $n$ users in the system. Also, we quantize the channel gain information to $q$ bits. Thus, the time required for reserving the frame for the best user for Qin & Berry's scheme is given by

$$T_r = \beta T_s' = \beta(\text{RTT} + (q + \log_2(n))T_b) = \beta(t' + q + \log_2(n))T_b. \tag{37}$$

Thus, our scheme is relatively more efficient than that of Qin & Berry when

$$\beta(t' + q + \log_2(n))T_b > (t' + kr)T_b \tag{38}$$

$$\text{i.e.,} \qquad \beta > \frac{t' + kr}{t' + q + \log_2(n)}. \tag{39}$$

Note that (39) applies to the case when $kr \geq q + \log_2(n)$, as $\beta \geq 1$. For the case when $kr < q + \log_2(n)$, our scheme is always better.

| $T_s$ in sec | RTT in $T_s, (t)$ | $T_c$ in $T_s, (p)$ |
|:---:|:---:|:---:|
| $10^{-9}$ | 3334 | 30000 |
| $10^{-8}$ | 334 | 3000 |
| $10^{-7}$ | 34 | 300 |

Table I

SIMULATION PARAMETERS

## VI. NUMERICAL RESULTS

In this section, we present the throughput or spectral efficiency achieved by the proposed CS-based reservation scheme for both analog and digital cases. We use maximum correlation technique for compressive sensing [8] as this is computationally much more efficient than $\ell_1$-minimization. Also, we use the following parameters in our simulation (summarized in Table I):

1) $n = 100$ users

2) $T_c = 30 \times 10^{-6} sec$

3) Distance between the BS and the users is $500 \ m$. Thus, the propagation delay between the users and BS (assuming speed of signal is $3 \times 10^8 m/sec$) is $500/(3 \times 10^8) = 1.6667 \times 10^{-6} sec$, which yields a RTT of $3.3334 \times 10^{-6} sec$.

4) Minimum slot duration $T_s$ supported by users' MAC device is $10^{-9} sec$, $10^{-8} sec$ and $10^{-7} sec$. Note that $T_s$ is the reciprocal of the maximum data rate supported by the MAC device, where $T_s = T_a$ for the analog case and $T_s = T_b$ for the digital case.

### A. Analog Case

In this subsection, we present numerical results for the proposed CS-based reservation scheme for the analog case. Based on the simulation data, in Fig. 3 - Fig. 5, we present the throughput or spectral efficiency versus $c$ which is related to $r$ by $r = cs \log(n/s)$, for different values of $s$. Also, we plot the spectral efficiency achieved by Qin & Berry's scheme and the maximum spectral efficiency that can be achieved (corresponding to zero reservation time).

As we see from (32), our scheme is more efficient for MAC devices that support higher data rate (smaller values of $T_a$) as this reduces $T_{CS}$ thereby reducing the total reservation time and eventually resulting in larger spectral efficiency. This very fact can be observed from Fig. 3 - Fig. 5 where all the parameters except $T_a$ are kept unchanged. When $T_a$ is relatively much smaller than RTT it is good to have high sparsity levels $s$ as reservation time is primarily dominated by RTT (see Fig. 3 & Fig. 4). From the figures, we can also note that for MAC devices having $T_a = 10^{-8} sec$ or less, the dominant part in the reservation phase is the RTT. Thus, there is hardly any decrease in the reservation time by reducing $T_a$ further and consequently there is negligible increase in the spectral efficiency as evident from Fig. 3 - Fig. 4.

However, it is not always a good idea to increase the value of $s$ (beyond a point) when $T_a$ is large, as is clearly evident from Fig. 5. This is because when $T_a$ is relatively not much smaller that RTT, the transmission time corresponding to $r$ real numbers is either at par with RTT or dominates the total reservation time.

### B. Digital Case

In this section, we present numerical results for the proposed CS-based reservation scheme for the digital case. Based on the simulation data, in Fig. 6 - Fig. 8, we present the throughput or spectral efficiency versus number of slots (or bits) used by CS per threshold, for different values of $k$. Also, we plot the spectral efficiencies achieved by Qin & Berry's scheme for $q = 4, 8$ and 16 bits and the maximum spectral efficiency that can be achieved (corresponding to zero reservation time).

As expected, trends similar to the analog case are observed in the digital case as evident from Fig. 6 - Fig. 8 where all the parameters except $T_b$ are kept unchanged. The scheme performs well for small values of $T_b$. However, the gain from reducing $T_b$ saturates once $T_{CS}$ becomes negligible in comparison to RTT. From reservation time perspective, the role played by $s$ in the analog case is played by $k$ in the digital case, and therefore increasing the number of thresholds (beyond a point) when $T_b$ is large is not a good idea as evident from Fig. 8. This is because when $T_b$ is relatively not much smaller that RTT, the transmission time corresponding to $kr$ bits is either at par with RTT or dominates the total reservation time. Note that the same applies to the number of quantization bits used in the Qin & berry's case, i.e. do not increase $q$ (beyond a point) when $T_b$ is relatively not much smaller that RTT.

Also, it is evident that the proposed scheme outperforms Qin & Berry's scheme in all cases considered in this paper.

### VII. Conclusions

In this paper, compressive sensing based opportunistic protocol for throughput improvement is proposed. Both analog and digital versions of the protocol were considered where the channel gains trans-mitted by the users were analog and digital, respectively. We have shown that the proposed protocol requires less time for reservation and so it enhances the performance of R-ALOHA, i.e. achieve better throughput than other R-ALOHA schemes proposed in literature. Using asymptotic analysis in the analog case, we show that the only hit on the throughput is due to reservation (i.e. it is asymptotically equivalent to the centralized scheme). Also, as the proposed scheme requires less reservation time, it can be seen as an enhancement for R-ALOHA schemes in fast fading environments.

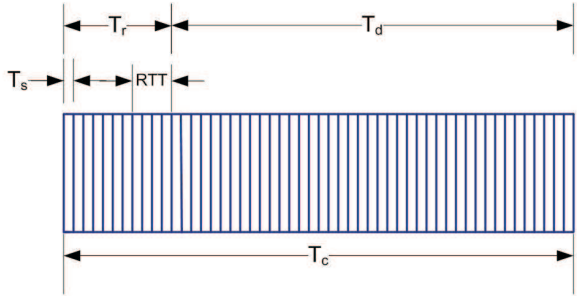

Figure 1.   An uplink frame of the proposed scheme

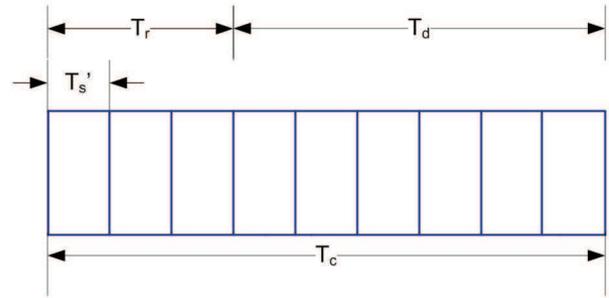

Figure 2.   An uplink frame of Qin & Berry's scheme

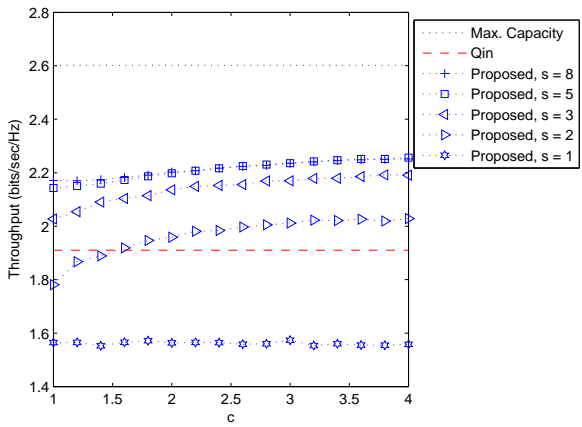

Figure 3.   Analog case: Throughput vs. $c$, $n = 100$, $T_a = 10^{-9} sec$, and $T_c = 30 \times 10^{-6} sec$ for different values of $s$.

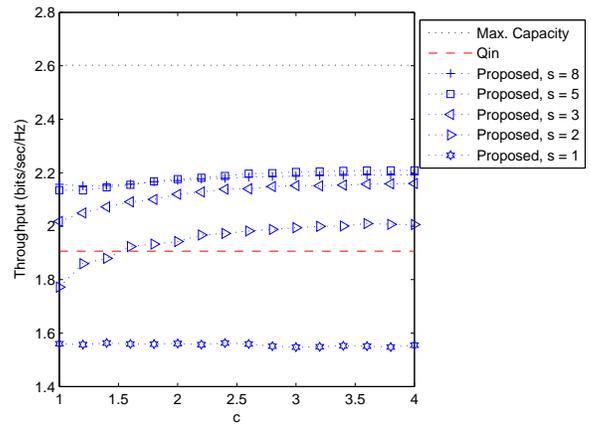

Figure 4.   Analog case: Throughput vs. $c$, $n = 100$, $T_a = 10^{-8} sec$, and $T_c = 30 \times 10^{-6} sec$ for different values of $s$.

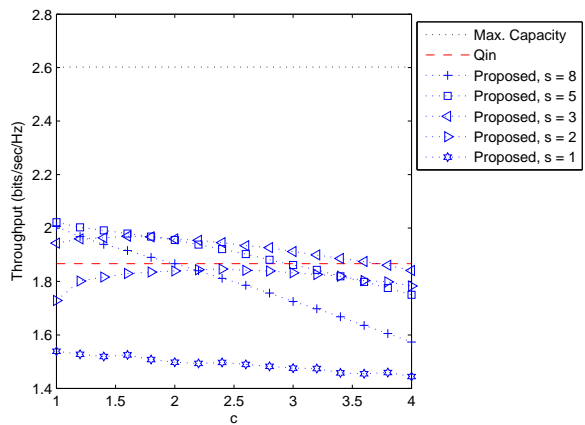

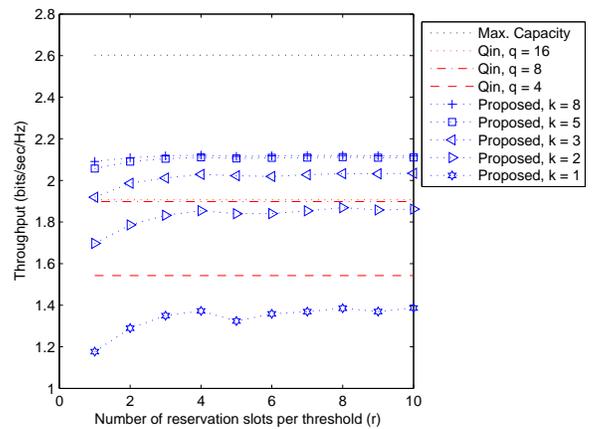

Figure 5. Analog case: Throughput vs. $c$, $n = 100$, $T_a = 10^{-7}sec$, and $T_c = 30 \times 10^{-6}sec$ for different values of $s$.

Figure 6. Digital case: Throughput vs. $r$, $n = 100$, $T_b = 10^{-9}sec$, $q = 4, 8$ and $16$ bits, and $T_c = 30 \times 10^{-6}sec$ for different values of $k$.

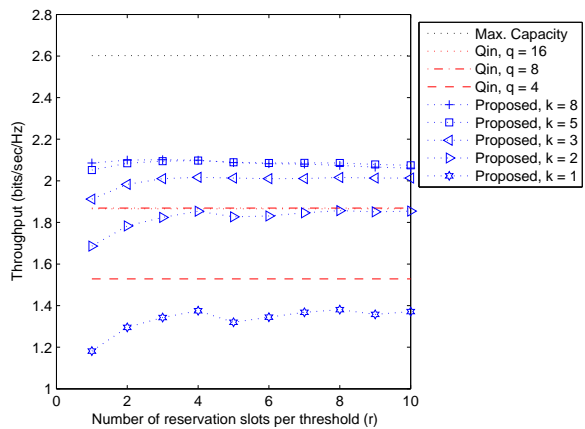

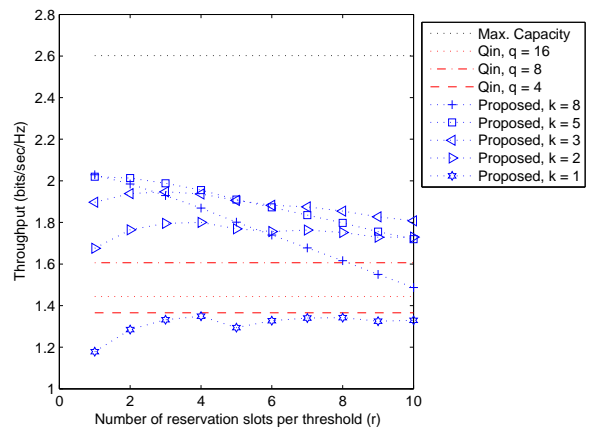

Figure 7. Digital case: Throughput vs. $r$, $n = 100$, $T_b = 10^{-8}sec$, $q = 4, 8$ and $16$ bits, and $T_c = 30 \times 10^{-6}sec$ for different values of $k$.

Figure 8. Digital case: Throughput vs. $r$, $n = 100$, $T_b = 10^{-7}sec$, $q = 4, 8$ and $16$ bits, and $T_c = 30 \times 10^{-6}sec$ for different values of $k$.